# Analysis of Repairable Systems Availability with Lindley Failure and Repair Behavior

Afshin Yaghoubi[1], Ph.D. candidate

Faculty of Mathematics and Computer Science, Amirkabir University of Technology, Tehran, Iran;

Email: afshin.y@aut.ac.ir

**Abstract**

Maintainability analysis is a cornerstone of reliability engineering. While the Markov approach is the classical analytical foundation, its reliance on the exponential distribution for failure and repair times is a major and often unrealistic limitation. This paper directly overcomes this critical constraint by investigating and modeling system maintainability using the more flexible and versatile Lindley distribution, which is represented via phase-type distributions. We first present a comprehensive maintainability analysis of a single-component system, deriving precise closed-form expressions for its time-dependent and steady-state availability, as well as the mean time to repair. The core methodology is then systematically generalized to analyze common series and parallel system configurations with n independent and identically distributed components. A dedicated numerical study compares the system performance under the Lindley and exponential distributions, conclusively demonstrating the significant and practical impact of non-exponential repair times on key reliability metrics. Our work provides a versatile and more widely applicable analytical framework for accurate maintainability assessment that successfully relaxes the restrictive exponential assumption, thereby offering greater realism in reliability modeling.

**Keywords:** Maintainability, Lindley distribution, Phase-type distribution, Availability.

## 1. Introduction

The Lindley distribution is a widely used statistical distribution in Bayesian inference, introduced by Lindley in 1958. This distribution is derived from a convex linear combination of an exponential distribution with scale parameter $\lambda$ and a gamma distribution with shape parameter 2 and scale parameter $\lambda$, with specified weights. The statistical properties of this distribution were thoroughly examined in [1], demonstrating that the Lindley distribution is more flexible than the exponential distribution in many aspects.

---

[1] Corresponding Author

Later, various generalizations of this distribution were introduced and applied in modeling lifetime data and, additionally, in the field of wind energy. For instance, references such as [2-5] can be consulted. Results in [2-3] indicate that this distribution outperforms major lifetime distributions like Weibull, log-normal, gamma, and exponential in modeling lifetime data. Furthermore, in the field of wind energy, where the Weibull distribution is the most common, the superiority of the Lindley distribution compared to Weibull is also demonstrated in [4-5].

The application of phase-type distributions in reliability is highly valuable, as it enables the analysis of a component's non-exponential failure time by modeling its passage through a series of intermediate phases, each with an exponential sojourn time.

Consider a system that transitions through several intermediate phases (states) before eventually reaching a failure (absorbing) state. The time to failure is a random variable with a phase-type distribution. A phase-type distribution is the distribution of the time to absorption in a continuous-time Markov chain with a finite number of states, comprising several transient states and one absorbing state. Phase-type distributions are extremely powerful because any distribution on positive real numbers can be approximated arbitrarily closely by a phase-type distribution [6]. These distributions are closed under operations such as summation, minimum, maximum, and convex combination of independent variables. This property is crucial for analyzing series and parallel systems [7]. Additionally, their computational simplicity, due to the ability to express their characteristics and statistical properties in matrix form, contributes to their importance. Many common distributions like exponential, Erlang, and hyper-exponential are special cases of phase-type distributions.

The [8], using phase-type distributions, transformed the reliability model of a two-component system with one repair facility into a Markov process with a specific matrix structure. Then, by leveraging this structure, they presented efficient numerical algorithms for computing important quantities such as the stationary state distribution and system time to failure. The [9] examined a single-component repairable system based on phase-type distributions with two repair patterns: perfect and minimal. By constructing the infinitesimal generator matrix of the Markov process and preserving the analytical properties of the exponential distribution, they provided explicit formulas for computing metrics such as availability, system operational time distribution, MTBF, and MTTR, and confirmed the model's applicability with numerical examples. The [10] reported on the dynamic assessment of multi-state non-repairable systems using phase-type modeling. They showed that the time the system spends in a subset of states can be modeled as a phase-type distribution. The [11] presented a model of an n-component system with one active unit and the others in cold standby, where failure and repair times follow discrete phase-type distributions. Using matrix methods, they computed reliability indices and also investigated the system state with infinite units. The [12] performed reliability analysis using the closure properties of phase-type distributions. Their proposed method was suitable for analyzing the reliability of structured systems and could generate the reliability functions of systems with independent components, not just point estimates of reliability. The [13] introduces an innovative generalization of phase-type distributions with multiple thresholds. This approach is suitable for modeling lifetime data in reliability analysis.

In repairable systems, the classical and widely used analytical approach is the Markov method, where failure and repair processes are modeled assuming exponentially distributed times. A substantial body of research has been built upon this framework (see, for example, works [14-16]). However, the core of these models remains reliant on the exponential distribution assumption, which is often restrictive in real-world scenarios.

Despite the wide application of phase-type distributions, no study to date has examined repairable systems with Lindley failure times within this framework. This paper, for the first time, presents the reliability analysis of repairable systems with Lindley failure times using the framework of phase-type distributions. This approach enables the calculation of long-term availability, mean time between failures, and reliability for industrial managers.

The paper is organized as follows: The next section introduces phase-type distributions and the representation of the Lindley distribution. The Lindley repairable single-component system model is presented in Section 3. Section 4 discusses the generalization to n-component Lindley systems. Section 5 presents numerical results, and the conclusion and future work are outlined in Section 6.

2. **Phase-Type Distribution**

If $T$ is a non-negative continuous random variable of phase-type, then its reliability function is defined as follows [10]:

$$R(t) = P(T > t) = \alpha e^{At} \mathbf{e}^T, \quad t \geq 0, \tag{1}$$

where $\mathbf{A}$ is an $m$ square matrix representing the transition intensity between transient states. The elements on the main diagonal of $\mathbf{A}$ are negative, and its other elements are non-negative. $\mathbf{e}^T$ is an $m \times 1$ column matrix with all entries equal to 1, and $\boldsymbol{\alpha}$ is a row probability vector $\boldsymbol{\alpha} = (\alpha_1, \dots, \alpha_m)$ whose components are non-negative and $\sum_{i=1}^m \alpha_i = 1$. Under these conditions, we say $T \sim \text{PH}(\boldsymbol{\alpha}, \mathbf{A})$, namely $T$ follows a phase-type distribution of order $m$, with transition matrix $\mathbf{A}$ and initial probability vector $\boldsymbol{\alpha}$.

It is known that the Lindley distribution can be expressed as $f_L(t; \lambda) = \alpha_1 f_E(t; \lambda) + \alpha_2 f_G(t; 2, \lambda)$ where the weights are given by $\alpha_1 = \frac{\lambda}{\lambda+1}$ and $\alpha_2 = 1 - \alpha_1$. Here, $f_E(t; \lambda) = \lambda e^{-\lambda t}$ and $f_G(t; 2, \lambda) = \lambda^2 t e^{-\lambda t}$.

This distribution can be interpreted as follows: a non-negative random variable $T$ follows an exponential distribution with probability $\alpha_1$, and a gamma distribution with shape parameter 2 with probability $\alpha_2$. For better understanding, Figure 1 illustrates its corresponding Markov chain.

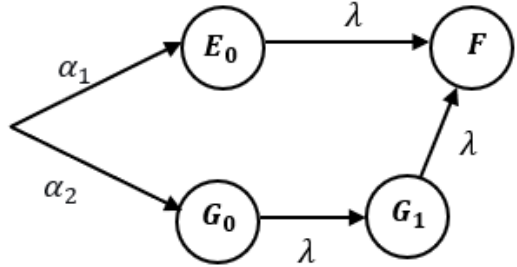

**Figure. 1.** Markov chain for the Lindley distribution as a phase-type representation

In Figure 1, there are 3 transient states $E$, $G_0$, and $G_1$, and one absorbing state $F$. State $E$ represents the exponential component, which is selected with probability $\alpha_1$ and transitions directly to the failure state $F$ at rate $\lambda$. States $G_0$, and $G_1$ represent the gamma component with shape parameter 2, which we model as a sequence of two exponential phases: first, $G_0$ is selected with probability $\alpha_2$ and transitions to $G_1$ at rate $\lambda$; then, from $G_1$, it transitions to the failure state $F$ at rate $\lambda$.

Note that in the gamma path, starting from state $G_1$ is never possible; therefore, its initial probability is zero.

To represent this distribution in phase-type form, the transition matrix $\mathbf{A}$ of order 3 and the initial probability vector $\boldsymbol{\alpha}$ are defined as follows:

$$\mathbf{A} = \begin{bmatrix} -\lambda & 0 & 0 \\ 0 & -\lambda & \lambda \\ 0 & 0 & -\lambda \end{bmatrix}, \boldsymbol{\alpha} = \left(\frac{\lambda}{\lambda+1}, \frac{1}{\lambda+1}, 0\right)$$

According to Equation (1), the reliability is given by:

$$R(t) = \left(\frac{\lambda}{\lambda+1}, \frac{1}{\lambda+1}, 0\right) \begin{bmatrix} e^{-\lambda t} & 0 & 0 \\ 0 & e^{-\lambda t} & \lambda t e^{-\lambda t} \\ 0 & 0 & e^{-\lambda t} \end{bmatrix} \begin{pmatrix} 1 \\ 1 \\ 1 \end{pmatrix} = \left(1 + \frac{\lambda t}{\lambda+1}\right) e^{-\lambda t}. \tag{2}$$

## 3. Analysis of a Single-Component Repairable System under the Lindley Distribution

Consider a single-component system. According to Figure 1, the state space of the chain is $S = \{E_0, G_0, G_1, F\}$. Now, the assumption of repairability is added to it. It is assumed that there is a repairman in the model who will repair the faulty component in case of system failure. If the system fails in the exponential path, repair is performed at rate $\mu \alpha_1$, and if it fails in the gamma path, repair is performed at rate $\mu \alpha_2$.

The starting states are also clearly defined in Figure 1. That is, $P_{E_0}(0) = \alpha_1$, $P_{G_0}(0) = \alpha_2$, and $P_{G_1}(0) = P_F(0) = 0$.

When the system enters the failed state $F$, a single repair process is initiated. The repair time is assumed to be exponentially distributed with rate $\mu$. Upon repair completion, the system is restored to an operational state according to the initial phase distribution of the Lindley model: it returns to state $E_0$ with probability $\alpha_1$ and to state $G_0$ with probability $\alpha_2$. This ensures the lifetime after repair preserves the Lindley distribution structure.

Accordingly, the transition rate matrix $\boldsymbol{Q}$ for the continuous-time Markov chain is given by:

$$\boldsymbol{Q} = \begin{bmatrix} -\lambda & 0 & 0 & \lambda \\ 0 & -\lambda & \lambda & 0 \\ 0 & 0 & -\lambda & \lambda \\ \mu\alpha_1 & \mu\alpha_2 & 0 & -\mu \end{bmatrix}$$

Note that in the last row, the transition rates from the failed state $F$ to the operational states $E_0$ and $G_0$ are $\mu\alpha_1$ and $\mu\alpha_2$, respectively. These are obtained by multiplying the overall repair rate $\mu$ by the probability of restarting in each corresponding operational phase ($\alpha_1$ or $\alpha_2$).

Now, the matrix differential equations (or Kolmogorov equations) can be computed using $\frac{d}{dt}P(t) = P(t)Q$ as follows:

$$\frac{d}{dt}P_{E_0}(t) = -\lambda P_{E_0}(t) + \mu\alpha_1 P_F(t),$$
$$\frac{d}{dt}P_{G_0}(t) = -\lambda P_{G_0}(t) + \mu\alpha_2 P_F(t),$$
$$\frac{d}{dt}P_{G_1}(t) = \lambda P_{G_0}(t) - \lambda P_{G_1}(t),$$
$$\frac{d}{dt}P_F(t) = \lambda P_{E_0}(t) + \lambda P_{G_1}(t) - \mu P_F(t). \tag{3}$$

### 3.1. Availability function

The most important factor in repairable systems is the availability function. Availability refers to the probability that the system is operational at time $t$.

In Equation (3), all the states above, except $P_F(t)$, indicate the system operational states. Therefore, to availability analysis, they must be determined. To solve them, one can take the Laplace transform of Equation (3).

After taking the Laplace transform, all probability states can be determined in terms of $P_F(s)$. Thus, the $P_F(s)$ is given by:

$$P_F(s) = \frac{\lambda^2(s + \lambda + 1)}{s[(\lambda + 1)s^2 + (\lambda + 1)(2\lambda + \mu)s + \lambda(\lambda + 1)(\lambda + \mu) + \lambda\mu]}. \tag{4}$$

Now, by performing the inverse Laplace transform on Equation (4), $P_F(t)$ can be easily obtained. Consequently, the availability function, denoted as $A(t; \lambda, \mu) = 1 - P_F(t)$, is derived.

Through straightforward mathematical calculations, $A(t; \lambda, \mu)$ can be expressed as:

$$A(t; \lambda, \mu) = \frac{\mu(\lambda + 2)}{\lambda(\lambda + 1) + \mu(\lambda + 2)} + \frac{\lambda(\lambda + 1)}{\lambda(\lambda + 1) + \mu(\lambda + 2)}\left(\cosh\left(\frac{\omega}{2}t\right) - M \sinh\left(\frac{\omega}{2}t\right)\right)e^{-(\lambda + 0.5\mu)t}, \tag{5}$$

where $\omega = \sqrt{\mu^2 - \frac{4\lambda\mu}{\lambda+1}}$, and $M = \frac{(\mu-2)\lambda^2 + 2(\mu-1)\lambda - \mu}{\omega(\lambda+1)^2}$.

It can be verified that as $\mu \to 0$, Equation (5) reduces to the Lindley reliability function given in Equation (2).

### 3.2. Steady-state availability

The steady-state (or long-run) availability, we say $A_\infty$, is obtained from the limit of the availability function as $t \to \infty$. In other words,

$$A_\infty = \lim_{t \to \infty} A(t; \lambda, \mu) = \frac{\mu(\lambda + 2)}{\lambda(\lambda + 1) + \mu(\lambda + 2)}. \qquad (6)$$

It can be shown that the long-run Lindley availability for any non-negative $\lambda$ and $\mu$ is always greater than the long-run exponential availability. We know that the steady-state availability for an exponential is given by $A_\infty^E = \frac{\mu}{\lambda+\mu}$. By adding the term $(\lambda + 2)$ to both the numerator and denominator, $A_\infty^E$ can be rewritten as $A_\infty^E = \frac{\mu(\lambda+2)}{(\lambda+\mu)(\lambda+2)}$. On the other hand, according to Equation (6), the long-run availability for Lindley can be expressed as $A_\infty^L = \frac{\mu(\lambda+2)}{(\lambda+\mu)(\lambda+2)-\lambda}$. For any non-negative $\lambda$ and $\mu$, it is evident that the inequality $0 < (\lambda + \mu)(\lambda + 2) - \lambda < (\lambda + \mu)(\lambda + 2)$ always holds. Therefore, $\frac{1}{(\lambda+\mu)(\lambda+2)-\lambda} > \frac{1}{(\lambda+\mu)(\lambda+2)}$. Now, multiplying both sides of the obtained inequality by the positive expression $\mu(\lambda + 2)$, we have: $A_\infty^L > A_\infty^E$.

### 3.3. Reliability function

When the assumption of repairability is removed, i.e., $\mu = 0$, (system components are non-repairable), then for reliability analysis, two key metrics are the reliability function and the mean time to failure (MTTF). To calculate the reliability function, it is sufficient to in Equation (5) set the repair rate to zero ($\mu = 0$) in the availability function. Therefore,

$$R(t; \lambda) = A(t; \lambda, \mu = 0) = (1 + \alpha_2 \lambda t)e^{-\lambda t} = \frac{1 + \lambda + \lambda t}{\lambda + 1} e^{-\lambda t}. \qquad (7)$$

This is precisely the survival function of the Lindley distribution, which was also evaluated in this study using the phase-type framework. After determining the reliability function, the mean time to failure (MTTF) can be calculated. This metric is obtained from $\int_0^\infty R(t; \lambda) \, dt = \frac{\lambda+2}{\lambda(\lambda+1)}$.

### 3.4. Mean time to repair

Another definition of steady-state availability is: $A_\infty = \frac{\text{MTTF}}{\text{MTTF}+\text{MTTR}}$, where MTTR is the mean time to repair a failed component. Now, MTTR can be easily calculated as $\text{MTTR} = \left(\frac{1-A_\infty}{A_\infty}\right)\text{MTTF} = \frac{1}{\mu}$.

### 4. Analysis of an *n*-Component Repairable System under the Lindley Distribution

Assume the system consists of two independent Lindley components. In this case, the state space increases to $4^2$, or 16 states. The 16 different states can be represented as ordered pairs in the Table 1:

**Table 1.** State space for a two-unit system with Lindley failure times.

| Stutes | First Unit | Second Unit | # of Failures |
|---|---|---|---|
| 1 | $E_0$ | $E_0$ | 0 |
| 2 | $E_0$ | $G_0$ | 0 |
| 3 | $E_0$ | $G_1$ | 0 |
| 4 | $E_0$ | $F$ | 1 |
| 5 | $G_0$ | $E_0$ | 0 |
| 6 | $G_0$ | $G_0$ | 0 |
| 7 | $G_0$ | $G_1$ | 0 |
| 8 | $G_0$ | $F$ | 1 |
| 9 | $G_1$ | $E_0$ | 0 |
| 10 | $G_1$ | $G_0$ | 0 |
| 11 | $G_1$ | $G_1$ | 0 |
| 12 | $G_1$ | $F$ | 1 |
| 13 | $F$ | $E_0$ | 1 |
| 14 | $F$ | $G_0$ | 1 |
| 15 | $F$ | $G_1$ | 1 |
| 16 | $F$ | $F$ | 2 |

The probability matrix for the table above is:

$$Q = \begin{bmatrix}
-2\lambda & 0 & 0 & 0 & 0 & 0 & 0 & 0 & 0 & 0 & 0 & 0 & 0 & \lambda & \lambda & 0 \\
0 & -2\lambda & \lambda & 0 & 0 & 0 & 0 & 0 & 0 & 0 & 0 & 0 & 0 & 0 & 0 & \lambda \\
0 & 0 & -\lambda & 0 & 0 & 0 & 0 & 0 & \lambda & 0 & 0 & 0 & 0 & 0 & 0 & 0 \\
0 & 0 & 0 & -2\lambda & 0 & 0 & 0 & 0 & 0 & 0 & 0 & 0 & \lambda & 0 & 0 & 0 \\
0 & 0 & 0 & 0 & -2\lambda & \lambda & 0 & 0 & 0 & 0 & 0 & 0 & 0 & 0 & 0 & \lambda \\
0 & 0 & 0 & 0 & 0 & -\lambda & 0 & 0 & \lambda & 0 & 0 & 0 & 0 & 0 & 0 & 0 \\
0 & 0 & 0 & 0 & 0 & 0 & -2\lambda & 0 & -2\lambda & \lambda & 0 & 0 & 0 & 0 & 0 & 0 \\
0 & 0 & 0 & 0 & 0 & 0 & 0 & -2\lambda & 0 & 0 & \lambda & 0 & 0 & 0 & 0 & 0 \\
0 & 0 & 0 & 0 & 0 & 0 & 0 & 0 & 0 & \lambda & 0 & 0 & 0 & 0 & 0 & 0 \\
0 & 0 & 0 & 0 & 0 & 0 & 0 & 0 & 0 & -(\lambda+\mu) & 0 & \lambda & 0 & 0 & 0 & \lambda \\
0 & 0 & 0 & 0 & 0 & 0 & 0 & 0 & 0 & 0 & -(\lambda+\mu) & \lambda & 0 & 0 & \lambda & 0 \\
\mu\alpha_1 & 0 & 0 & \mu\alpha_2 & 0 & 0 & 0 & 0 & 0 & 0 & 0 & -\mu & 0 & 0 & 0 & \lambda \\
0 & \mu\alpha_1 & 0 & 0 & \mu\alpha_2 & 0 & 0 & 0 & 0 & 0 & 0 & 0 & -(\lambda+\mu) & 0 & 0 & \lambda \\
0 & 0 & \mu\alpha_1 & 0 & 0 & \mu\alpha_2 & 0 & 0 & 0 & 0 & 0 & 0 & 0 & -(\lambda+\mu) & 0 & \lambda \\
0 & 0 & 0 & 0 & 0 & 0 & \mu\alpha_1 & \mu\alpha_2 & 0 & 0 & 0 & 0 & 0 & 0 & -(\lambda+\mu) & \lambda \\
0 & 0 & 0 & 0 & 0 & 0 & 0 & 0 & \mu\alpha_1 & \mu\alpha_2 & \mu\alpha_1 & \mu\alpha_2 & 0 & 0 & 0 & -2\mu
\end{bmatrix}$$

The differential equations governing the system are determined using $\frac{d}{dt}\mathbf{P}(t) = \mathbf{P}(t)\mathbf{Q}$. Obtaining an explicit parametric solution for these equations to derive the availability function is somewhat challenging. However, the important metric of long-run availability can be calculated by solving the system of algebraic equations resulting from the steady-state condition $\frac{d}{dt}\mathbf{P}(\infty) = \mathbf{0}$, i.e., $\mathbf{P}(\infty)\mathbf{Q} = \mathbf{0}$.

Given the complete independence of the component failure and repair processes, and assuming a separate repairman is available for each component, the steady-state availability of an $n$-component system in series and parallel configurations can be directly computed from the single-component availability. If $A_\infty^{(i)}$ represents the long-term availability of the $i$-th component, as obtained from the previous analysis, then the steady-state availability for an $n$-component system in series and parallel configurations is respectively given by $A_\infty^{\text{Series}} = \prod_{i=1}^n A_\infty^{(i)}$ and $A_\infty^{\text{Parallel}} = 1 - \prod_{i=1}^n \left(1 - A_\infty^{(i)}\right)$. The computational formulas for $A_\infty^{\text{Series}}$ and $A_\infty^{\text{Parallel}}$ are derived from the following relations:

$$A_\infty^{\text{Series}} = \prod_{i=1}^n \frac{\mu_i(\lambda_i + 2)}{\lambda_i(\lambda_i + 1) + \mu_i(\lambda_i + 2)}, \tag{8}$$

$$A_\infty^{\text{Parallel}} = 1 - \prod_{i=1}^n \left(\frac{\lambda_i(\lambda_i + 1)}{\lambda_i(\lambda_i + 1) + \mu_i(\lambda_i + 2)}\right). \tag{9}$$

Equations (8) and (9) provide the long-run availability for systems configured in series or in parallel, each consisting of $n$ independent components with Lindley failure and repair rates. As many industrial production lines follow such configurations, these formulas are highly applicable for production and maintenance management.

## 5. Numerical Example

In this section, we examine a numerical example from previous work that considered the repairability hypothesis based on the exponential distribution. The [17], analyzes a biomass-fueled combined cooling, heating, and power (CCHP) system consisting of three main subsystems arranged in series: 1) a gasification system (G) that converts biomass into syngas, 2) an internal combustion engine (ICE) that generates electricity using the syngas, and 3) an absorption chiller (AC) that produces chilled water for cooling. The system is considered fully available (State 1) only when all three subsystems (G, ICE, and AC) are functioning simultaneously.

The information related to the failure rates and repair rates of the system from [17] is listed in Table 2:

**TABLE 2.** The failure and repair rates of CCHP system [17]

| Subsystem | $\lambda$, per day | $\mu$, per day |
|---|---|---|
| G | 0.004 | 0.03 |
| ICE | 0.002 | 0.08 |
| AC | 0.002 | 0.08 |

### 5.1. Reliability analysis

According to Table 2, the reliability of each individual subsystem, with repair and without repair ($\mu = 0$), was examined. Figure 2 illustrates the results.

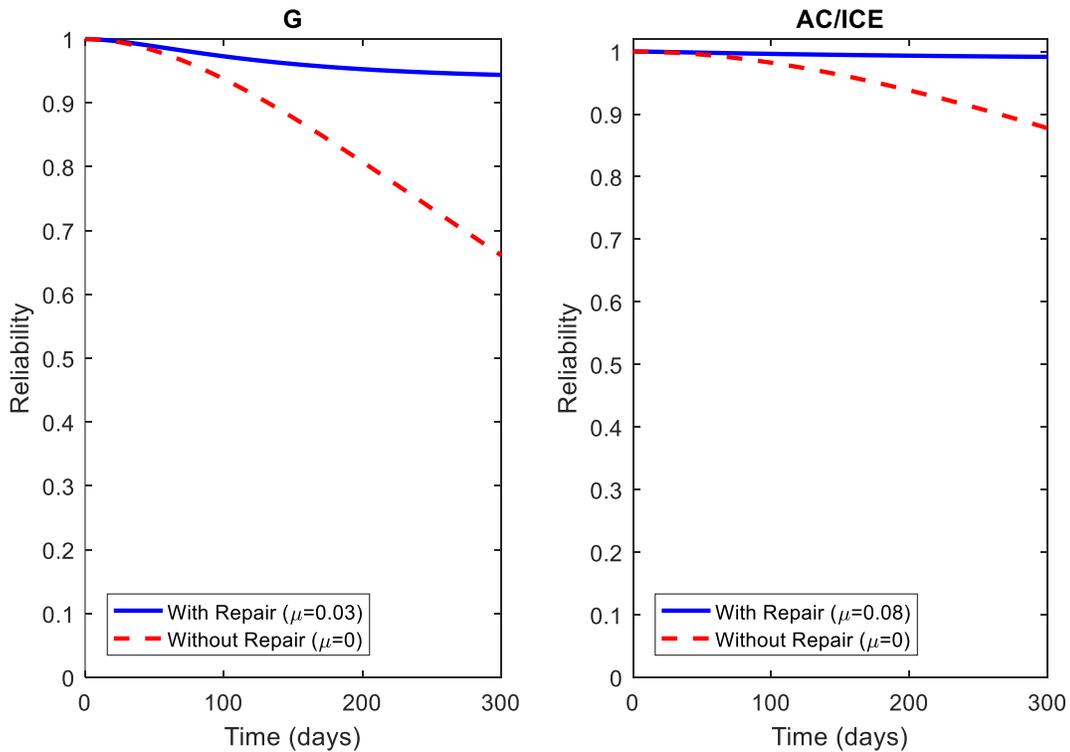

**Figure 2.** Reliability plots for CCHP subsystems with and without repair

Figure 2 compares the reliability of each subsystem with and without repair. The solid curves (with repair) exhibit an initial decline before stabilizing at a steady-state value, reflecting the balance between failure and repair processes. In contrast, the dashed curves (without repair) show monotonic exponential decay. The higher reliability of ICE and AC subsystems compared to the G subsystem is attributed to their lower failure rates and higher repair rates. The identical reliability profiles of ICE and AC result from their identical failure and repair parameters.

Here, we are interested in comparing the time-dependent availability of individual subsystems under the Lindley and exponential failure time assumptions. For this purpose, we use Equation (5) to compute $A_L(t; \lambda, \mu)$ under the Lindley distribution and the classical analytical formula for single-component exponential systems $A_E(t; \lambda, \mu) = \frac{\mu}{\lambda+\mu} + \frac{\lambda}{\lambda+\mu} e^{-(\lambda+\mu)t}$ for the exponential model.

Figure 3 show the results of this comparison for the G subsystem and the AC/ICE subsystems (which share identical parameters), respectively, using the parameters from Table 2.

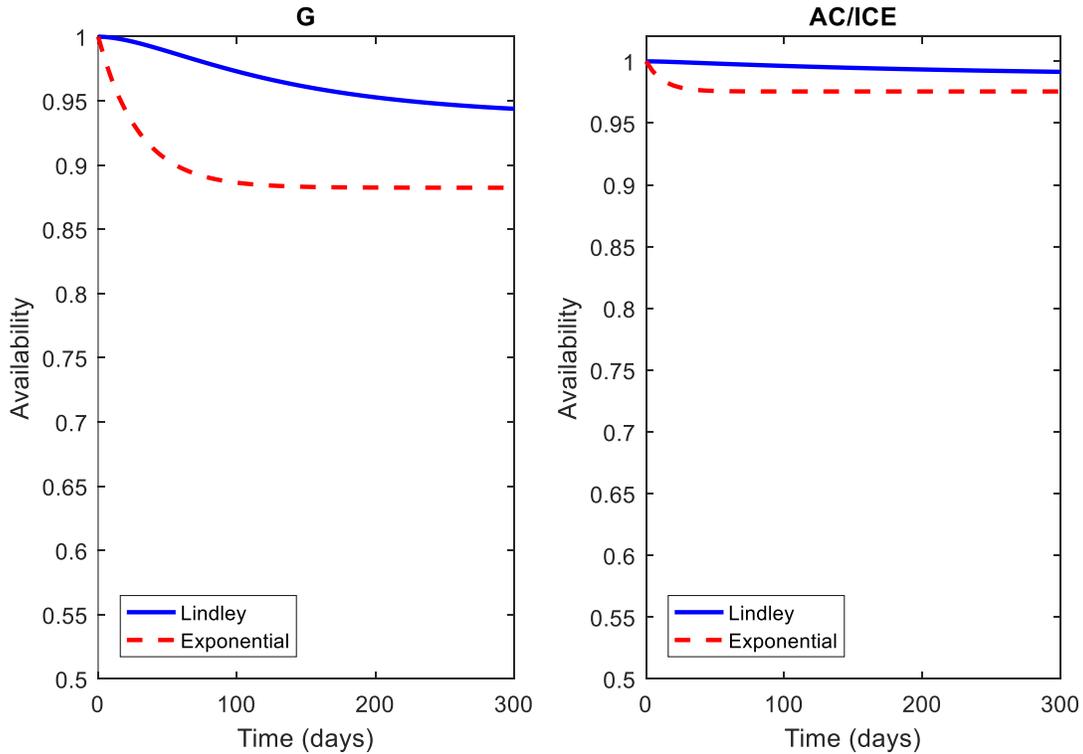

**Figure 3.** Availability plots for CCHP subsystems- Lindley vs. Exponential

It is observed that throughout the entire time horizon, the Lindley availability curve $A_L(t; \lambda, \mu)$ lies systematically higher than the exponential curve $A_E(t; \lambda, \mu)$. The more complex transient structure of the Lindley model (due to its phase-type representation) compared to the exponential model is evident in the shape of these curves.

These results emphasize that the exponential model can provide a conservative (pessimistic) estimate of the operational availability of components across all time horizons. Adopting the more realistic Lindley model, which can capture non-exponential failure time behavior, appears particularly essential for components with lower reliability (such as the G subsystem in this study) for more accurate maintenance and repair planning.

### 5.2. Comparison of $A_\infty$ under Lindley and Exponential distributions

This section presents a comparative analysis of the $A_\infty$ for each subsystem, evaluated under both the Lindley and exponential failure time distributions. The comparison aims to highlight the impact of distributional assumptions on $A_\infty$ estimates.

For a repairable system with failure rate $\lambda$ and repair rate $\mu$, the steady-state availability under the exponential distribution is given by the classical formula: $A_\infty^E = \frac{\mu}{\lambda+\mu}$, and under the Lindley distribution, as derived in Equation (6), the steady-state availability takes the form: $A_\infty^L = \frac{\mu(\lambda+2)}{\lambda(\lambda+1)+\mu(\lambda+2)}$. Using the failure and repair rates listed in Table 2, the long-term availabilities for each subsystem under both distributions are computed and summarized in Table 3.

**TABLE 3.** $A_\infty$ under exponential and Lindley assumptions

| Subsystem | $A_\infty^E$ | $A_\infty^L$ |
|---|---|---|
| G | 0.8824 | 0.9374 |
| AC/ICE | 0.9756 | 0.9876 |

The results in Table 3 clearly demonstrate that the Lindley distribution consistently yields higher long-run availability estimates than the exponential distribution for all subsystems. This upward shift stems from the Lindley distribution's ability to model more realistic failure behaviors—specifically, its representation of a mixture of immediate and delayed failure paths—which effectively reduces the apparent impact of the failure rate on long-run availability.

Since the system is in series, the overall system availability ($A_\infty^{\text{Series}}$) can be calculated using Equation (8). The results are presented in Table 4.

**TABLE 4.** $A_\infty$ under exponential and Lindley assumptions

| System | $A_\infty^{\text{Series}}$ | Relative Increase |
|---|---|---|
| Exponential | 0.8398 | - |
| Lindley | 0.9143 | 7.5% |

Table 4 presents the overall steady-state availability of the series-configured CCHP system under both exponential and Lindley failure-time assumptions. The Lindley distribution yields a system-level availability of 0.9143, which is 7.5% higher than the exponential-based estimate of 0.8398. This consistent upward shift reaffirms that modeling failure times with the more flexible Lindley distribution leads to more optimistic—and arguably more realistic—availability projections for the entire system.

### 5.3. Effect of failure rate

The sensitivity of the steady-state availability to changes in the failure rate $\lambda$ is quantified by its partial derivative. Differentiating Equation (6) with respect to $\lambda$ yields:

$$\frac{\partial A_\infty^L}{\partial \lambda} = -\frac{\mu(\lambda^2 + 4\lambda + 2)}{[\lambda(\lambda+1) + \mu(\lambda+2)]^2}. \tag{10}$$

The Equation (10) for each $\lambda$ and $\mu$ positive is always negative. In limiting behavior: $\lim_{\lambda \to 0} A_\infty^L = 1$, and $\lim_{\lambda \to \infty} A_\infty^L = 0$. This confirms that $A_\infty^L$ is a strictly decreasing function of $\lambda$.

### 5.4. Effect of repair rate

Similarly, the sensitivity to the repair rate $\mu$ is found by differentiating Equation (6) with respect to $\mu$:

$$\frac{\partial A_\infty^L}{\partial \mu} = \frac{\lambda(\lambda+1)(\lambda+2)}{[\lambda(\lambda+1) + \mu(\lambda+2)]^2}. \tag{11}$$

For all $\lambda$ and $\mu > 0$, the derivative in Equation (11) is always positive. In limiting behavior: $\lim_{\mu \to 0} A_\infty^L = 0$, and $\lim_{\mu \to \infty} A_\infty^L = 1$. This confirms that $A_\infty^L$ is a strictly increasing function of $\mu$.

We analyze the effect of varying the $\lambda$ around its nominal value for each subsystem while keeping its $\mu$ fixed. For the Gasification (G) Subsystem (Nominal: $\lambda = 0.004$, $\mu = 0.03$), and the ICE/AC Subsystems (Nominal: $\lambda = 0.002$, $\mu = 0.08$), Tables 5 and 6 show how $A_\infty^L$ for these subsystem changes with $\lambda$, respectively.

**TABLE 5.** Sensitivity of $A_\infty^L$ to $\lambda$ for the G Subsystem ($\mu = 0.03$ fixed)

| $\lambda$, per day | $A_\infty^L$ | $\frac{\partial A_\infty^L}{\partial \lambda}$ |
|---|---|---|
| 0.002 | 0.9677 | -15.5 |
| 0.004 (Nominal) | 0.9374 | -14.7 |
| 0.006 | 0.9083 | -13.9 |
| 0.008 | 0.8802 | -13.1 |

**TABLE 6.** Sensitivity of $A_\infty^L$ to $\lambda$ for the ICE/AC Subsystems ($\mu = 0.08$ fixed)

| $\lambda$, per day | $A_\infty^L$ | $\frac{\partial A_\infty^L}{\partial \lambda}$ |
|---|---|---|
| 0.001 | 0.9938 | -6.21 |
| 0.002 (Nominal) | 0.9876 | -6.11 |
| 0.003 | 0.9809 | -6.01 |
| 0.004 | 0.9738 | -5.91 |

The results demonstrate that $A_\infty^L$ is a decreasing function of $\lambda$, as expected. The sensitivity, measured by the partial derivative, is negative, and its magnitude increases with $\lambda$. The impact of changing $\lambda$ is more significant for the G subsystem. For instance, increasing $\lambda$ by 50% (from 0.004 to 0.006) reduces its availability from 0.9374 to 0.9083, a reduction of approximately 2.9 percentage points. In contrast, a similar 50% increase for the AC/ICE subsystems (from 0.002 to 0.003) reduces availability from 0.9876 to 0.9809, a smaller reduction of about 0.7 percentage points. This indicates that reliability improvement efforts (reducing $\lambda$) would yield greater benefits if focused on the G subsystem, which has a lower baseline availability.

We now analyze the effect of varying the $\mu$ around its nominal value for each subsystem while keeping its failure rate fixed. For the Gasification (G) Subsystem (Nominal: $\lambda = 0.004, \mu = 0.03$), and for the ICE/AC Subsystems (Nominal: $\lambda = 0.002, \mu = 0.08$), Tables 7 and 8 show how $A_\infty^L$ for these subsystem changes with $\mu$, respectively.

**TABLE 7.** Sensitivity of $A_\infty^L$ to $\mu$ for the G Subsystem ($\lambda = 0.004$ fixed)

| $\mu$, per day | $A_\infty^L$ | $\frac{\partial A_\infty^L}{\partial \mu}$ |
|---|---|---|
| 0.015 | 0.8820 | 3.91 |
| 0.030 (Nominal) | 0.9374 | 1.96 |
| 0.045 | 0.9605 | 1.17 |
| 0.060 | 0.9722 | 0.80 |

**TABLE 8.** Sensitivity of $A_\infty^L$ to $\mu$ for the ICE/AC Subsystem ($\lambda = 0.002$ fixed)

| $\mu$, per day | $A_\infty^L$ | $\frac{\partial A_\infty^L}{\partial \mu}$ |
|---|---|---|
| 0.04 | 0.9756 | 0.31 |
| 0.08 (Nominal) | 0.9876 | 0.15 |
| 0.12 | 0.9922 | 0.10 |
| 0.16 | 0.9945 | 0.07 |

The results confirm that $A_\infty^L$ is an increasing function of $\mu$, as expected. The sensitivity (partial derivative) is positive but decreases as $\mu$ increases, indicating diminishing returns on investment in repair speed. Improving the $\mu$ has a more substantial absolute impact on the G subsystem. For example, doubling $\mu$ from 0.03 to 0.06 increases its availability from 0.9374 to 0.9722, an improvement of approximately 3.5 percentage points. In contrast, doubling the repair rate for the AC/ICE subsystems (from 0.08 to 0.16) increases availability from 0.9876 to 0.9945, yielding a much smaller improvement of only about 0.7 percentage points, given their already high baseline availability. This suggests that maintenance resources aimed at speeding up repairs are more effectively allocated to the G subsystem.

## 6. Conclusuon

This paper introduced a novel phase-type framework for analyzing repairable systems with Lindley-distributed failure times, overcoming the exponential limitation of classical Markov models. We derived closed-form solutions for key metrics (availability, MTTF, MTTR) for single and multi-component (series/parallel) systems. The numerical study on a CCHP system showed that the Lindley model predicts 8.87% higher system availability than the exponential model. Sensitivity analysis identified the Gasification subsystem as the critical bottleneck, proving that maintenance efforts focused there yield the greatest improvement. In summary, this work provides a more flexible and realistic analytical tool for reliability engineering. Future work can explore dependent components, non-exponential repair times, and optimal maintenance policies within this framework.

[6] Bladt, M., & Nielsen, B. F. (2017). Phase-type distributions for insurance and finance. Wiley.

[7] Assaf, D., & Levikson, B. (1982). Closure of phase type distributions under operations arising in reliability theory. The Annals of Probability, 10(1), 265–269.

[8] Neuts, M. F., & Meier, K. S. (1981). On the use of phase type distributions in reliability modelling of systems with two components. *Operations-Research-Spektrum, 2*(4), 227–234.

[9] Peng, D., Fang, L., & Tong, C. (2013, July). A multi-state reliability analysis of single-component repairable system based on phase-type distribution. In *2013 International Conference on Management Science and Engineering 20th Annual Conference Proceedings* (pp. 496-501). IEEE.

[10] Eryılmaz, S. (2015). Dynamic assessment of multi-state systems using phase-type modeling. Reliability Engineering & System Safety, 143, 152–158.

[11] Ruiz-Castro, J. E., Pérez-Ocón, R., & Fernández-Villodre, G. (2008). Modelling a reliability system governed by discrete phase-type distributions. *Reliability Engineering & System Safety*, *93*(11), 1650-1657.

[12] Alkaff, A., & Qomarudin, M. N. (2020). Modeling and analysis of system reliability using phase-type distribution closure properties. *Applied Stochastic Models in Business and Industry*, *36*(4), 548-569.

[13] Ruiz-Castro, J. E., Acal, C., & Roldán, J. B. (2023). An approach to non-homogenous phase-type distributions through multiple cut-points. *Quality Engineering*, *35*(4), 619-638.

[14] Liang, Q., Yang, Y., Zhang, H., Peng, C., & Lu, J. (2022). Analysis of simplification in Markov state-based models for reliability assessment of complex safety systems. Reliability Engineering & System Safety, 221, 108373.

[15] Kumar, P., Singh, L. K., & Kumar, C. (2019). An optimized technique for reliability analysis of safety-critical systems: A case study of nuclear power plant. Quality and Reliability Engineering International, 35(1), 461-469.

[16] Son, K. S., Kim, D. H., Kim, C. H., & Kang, H. G. (2016). Study on the systematic approach of Markov modeling for dependability analysis of complex fault-tolerant features with voting logics. Reliability Engineering & System Safety, 150, 44-57.
16